\documentclass[Physsubmission, Phys]{SciPost}
\usepackage{amsmath}
\usepackage{subcaption}
\usepackage{overpic}

\hypersetup{
    colorlinks,
    linkcolor={red!50!black},
    citecolor={blue!50!black},
    urlcolor={blue!80!black}
}

\usepackage[bitstream-charter]{mathdesign}
\DeclareSymbolFont{usualmathcal}{OMS}{cmsy}{m}{n}
\DeclareSymbolFontAlphabet{\mathcal}{usualmathcal}

\newcommand{\Lambdabar}{\bar{\Lambda}}
\newcommand{\jpsi}{J/\psi}
\newcommand{\comment}[1]{}
\begin{document}

\begin{center}{\Large \textbf{
Hyperon Physics at BESIII\\
}}\end{center}

\begin{center}
Viktor Thorén (on behalf of the BESIII collaboration)\textsuperscript{1}*,
\end{center}

\begin{center}
{\bf 1} Uppsala University, Box 516, SE-75120, Uppsala, Sweden \\
* viktor.thoren@physics.uu.se
\end{center}

\begin{center}
\today
\end{center}


\definecolor{palegray}{gray}{0.95}
\begin{center}
\colorbox{palegray}{
  \begin{tabular}{rr}
  \begin{minipage}{0.1\textwidth}
    \includegraphics[width=23mm]{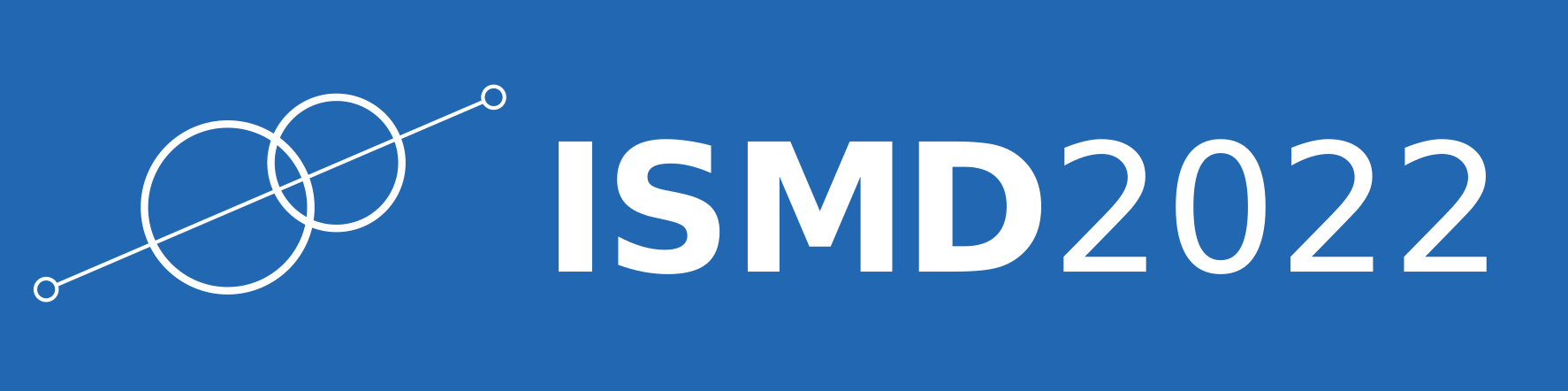}
  \end{minipage}
  &
  \begin{minipage}{0.8\textwidth}
    \begin{center}
    {\it 51st International Symposium on Multiparticle Dynamics (ISMD2022)}\\ 
    {\it Pitlochry, Scottish Highlands, 1-5 August 2022} \\
    \doi{10.21468/SciPostPhysProc.?}\\
    \end{center}
  \end{minipage}
\end{tabular}
}
\end{center}

\section*{Abstract}
{\bf
Spin polarization and entanglement are utilized by the BESIII experiment to learn more about the production and decay properties of hyperon-antihyperon pairs in a series of recent analyses of its unprecedented datasets at the $\jpsi$ and $\psi'$ resonances. This has lead to the observation of significant transverse polarisation in decays of $\jpsi$ or $\psi'$ into $\Lambda \Lambdabar$, $\Sigma \bar{\Sigma}$, $\Xi \bar{\Xi}$, and $\Omega \bar{\Omega}$. Because of the non-zero polarization, the decay parameters for the most common hadronic weak decays of both hyperons and antihyperons could be determined independently for the first time. Comparing these to each other yields precise tests of direct, $\Delta S = 1$ CP-violation that complement measurements of kaon decays.
}

\section{Introduction}
Decays of hyperon-antihyperon pairs produced in $e^+e^-$-annihilation are excellent laboratories for precise tests of fundamental discrete symmetries such as CP. Recently, it was shown that a long-standing formalism used to describe the production of hyperon-antihyperon pairs via one photon exchange~\cite{Dubnickova:1992ii, GAKH2006169, PhysRevD.75.074026, Faldt:2013gka, Faldt:2016qee} also holds for intermediate vector resonances such as $J/\psi$ and $\psi'$~\cite{Faldt:2017kgy}. This opened up the possibility of exploiting the large datasets of $10^{10}$ $J/\psi$ and $3\times10^9$ $\psi'$ events collected by the BESIII experiment to perform detailed analyses of multidimensional angular distributions to extract polarizations and spin correlations in order to determine the production and decay properties. Applying the same method to hyperon-antihyperon pairs produced in one-photon exchange is challenging due to the limited size of available data samples, but can provide important insights into the hitherto relatively unexplored internal structure of hyperons.

\section{Direct CP-violation in Hyperon Decays}
In order to observe CP-violation in a given transition, there needs to be at least two interfering amplitudes with different CP-odd phases. Direct, $\Delta S = 1$ CP-violation was first seen in decays of kaons into two pions~\cite{PhysRevD.67.012005, BATLEY200297, PhysRevD.83.092001} where two isopsin transitions $\mathcal{A}_{\Delta I = 1/2}$ and $\mathcal{A}_{\Delta I = 3/2}$ contribute. The ratio of the partial decay widths of $K_L$ and $K_S$ gives the size of the CP-violating contribution
\begin{align}
        \frac{{\cal A}(K_L\to\pi^0\pi^0)}{{\cal A}(K_S\to\pi^0\pi^0)}:= \epsilon-2\epsilon' \, \text{and} \, 
        \frac{{\cal A}(K_L\to\pi^+\pi^-)}{{\cal A}(K_S\to\pi^+\pi^-)}:= \epsilon+\epsilon'.
\end{align}
The standard model (SM) mechanism for CP-violation, originating from the CKM mixing matrix via the so-called penguin diagrams, see \textit{e.g.} Ref.~\cite{Gisbert2018}, can successfully explain the observed values of $\epsilon$ and $\epsilon'$.

The most common decays of ground state hyperons are $\Delta S = 1$  weak decays into a baryon and a meson. Unlike the kaons, hyperons carry spin and therefore these decays receive parity-even $p$-wave and parity-odd $s$-wave contributions with amplitudes $P$ and $S$, respectively, for both $\Delta I = 1/2$ and $\Delta I = 3/2$. The full decay amplitude can be expressed as 
\begin{align}
    \mathcal{A} = S + P\sigma\cdot \hat{n},
\end{align}
where $\hat{n}$ is a unit vector along the final-state baryon momentum in the parent momentum rest frame. The contribution from the $\Delta I = 3/2$ transition is small and including the $\Delta I = 1/2$ transition is sufficient for a first order treatment. The $p$- and $s$-wave contributions can be expressed in terms of strong phases $\delta$ and CP-odd weak phases $\xi$ as
\begin{align}
    S &= |S| \text{exp}(i \xi_S) \text{exp}(i\delta_S),  \\
    P &= |P| \text{exp}(i \xi_P) \text{exp}(i\delta_P).
\end{align}
CP-violation occurs if the weak phase difference $|\xi_S-\xi_P|$ is non-zero. One can define  two parameters $\alpha$ and $\beta$ in terms of the interference of the two amplitudes to the describe the decay process~\cite{PhysRev.108.1645}
\begin{align}
        \alpha &= \frac{2 \text{Re}(S^*P)}{|S|^2+|P|^2} \\
        \beta &= \frac{2 \text{Im}(S^*P)}{|S|^2+|P|^2} = \sqrt{1-\alpha^2}\sin\phi,
\end{align}
where $\phi$ is introduced because it has a more straightforward experimental interpretation. If the polarization of the decaying hyperon and the angular distribution of the final state baryon can be measured, the parameter $\alpha$ can be determined in any one-step decay, \emph{e.g.} $\Lambda \to p \pi^-$. To measure $\phi$, one needs a sequential decay process, such as $\Xi \to \Lambda\pi^-, \, \Lambda \to p\pi^-$ where the polarization of the parent and decay product hyperons can be compared.

Assuming that CP is conserved, the decay parameters for baryons and antibaryons should be equal up to a relative sign, $\alpha = - \bar{\alpha}$, $\beta = - \bar{\beta}$, $\phi = - \bar{\phi}$. One can define two independent tests of CP-symmetry
\begin{align}
A_{CP} &= \frac{\alpha+ \bar{\alpha}}{\alpha-\bar{\alpha}} \\
B_{CP} &= \frac{\beta + \bar{\beta}}{\alpha-\bar{\alpha}} = (\xi_P - \xi_S),
\end{align} 
where a non-zero value of $A_{CP}$ or $B_{CP}$ would indicate CP-violation. In the SM, the largest CP-violating effects are expected in the decays $\Xi^- \to \Lambda \pi^-$ and $\Lambda \to p \pi^-$ with predictions for $A_{CP}$ on the order of $10^{-5}$~\cite{Tandean:2002vy}. If a larger value were to be measured, it could be used to predict beyond SM contributions to CP-violation in the kaon sector~\cite{Tandean:2003fr}.

On the experimental side, there have been no independent measurements of the decay parameters of hyperons and antihyperons until now. The most precise result comes from the HyperCP experiment that measured the combination $A_{CP}^{\Xi} + A_{CP}^{\Lambda} = 0(5)(5)\times10^{-4}$~\cite{HyperCP:2004zvh}.  Furthermore, the PS185 experiment studied the product of $\alpha_\Lambda$ and the $\Lambda$ polarization to determine $A_{CP}^{\Lambda} = 0.013 \pm0.021$~\cite{PhysRevC.54.1877}.

\section{Baryon-Antibaryon Production in $e^+e^-$ Annihilation}
 Even if the initial state is unpolarized, any type of baryon-antibaryon pair produced in $e^+e^-$-annihilation can be polarized in the direction perpendicular to the production plane. The production mechanism dictates the degree of polarization, if any. The level of complexity required to describe the production process depends on the spin of the baryons. The production and decay of a spin-3/2 baryon-antibaryon pair is described by four complex form factors, while two  are sufficient for a spin-1/2 baryon-antibaryon pair. In the following we will consider the spin-1/2 case in more detail. Regardless of whether it occurs through the exchange of a single photon or via an intermediate vector resonance, the production of a spin-1/2 baryon antibaryon pair is described by two complex form factors $G_E^\psi$ and $G_M^\psi$~\cite{Faldt:2017kgy}. These can be related to two observable parameters associated with the baryon scattering angle and polarization: $\alpha_\psi$ which is connected to the ratio of form factors $R= | G_E^\psi / G_M^\psi|$, and $\Delta \Phi$ which is the relative phase between the form factors. The transverse polarization of the baryon and antibaryon is non-zero if $\Delta \Phi$ is non-zero. If and only if this is the case, it is possible to determine the decay parameters of hyperons and antihyperons simultaneously and independently of each other.

The authors of Ref.~\cite{ModApproach} have developed a modular approach that can be used to describe the production and decay of baryon-antibaryon pairs with any combinations of spin-1/2 or spin-3/2. The spin polarization and correlations are encoded in the spin-density matrix which, for spin-1/2, is given by
\begin{align}
    \rho_{B\bar{B}} = \sum C^{1/2}_{\mu\nu} \sigma_\mu^B\otimes \sigma_\nu^{\bar{B}},
\end{align}
where $\sigma_\mu^B(\sigma_\nu^{\bar{B}})$ are the Pauli matrices in the (anti)baryon helicity frame. Information on the spin correlations $C_{ij}(\theta)$ and polarization $P_y(\theta)$ is found in the coefficient matrix $C_{\mu\nu}$. For a spin-1/2 baryon-antibaryon pair, it is given by
\begin{align}
\begin{split}
    C^{1/2}_{\mu\nu} &= 3 \frac{1+\alpha_\psi \cos^2\theta}{3 + \alpha_\psi}\begin{pmatrix} 1  & 0 & P_y& 0 \\ 0 & C_{xx} & 0& C_{xz} \\ -P_y & 0 & C_{yy}& 0 \\ 0 & -C_{xz} & 0 & C_{zz}  \end{pmatrix} \\
    &=  \frac{3}{3 + \alpha_\psi}\begin{pmatrix} 1 + \alpha_\psi \cos^2 \theta  & 0 & \beta_\psi \sin \theta \cos \theta & 0 \\ 0 & \sin^2 \theta & 0& \gamma_\psi \sin \theta \cos \theta  \\ -\beta_\psi \sin \theta \cos \theta  & 0 & \alpha_\psi \sin^2\theta & 0 \\ 0 & -\gamma_\psi \sin \theta \cos \theta & 0 & -\alpha_\psi - \cos^2\theta  \end{pmatrix},
    \end{split}
\end{align}
where $\beta_\psi = \sqrt{1-\alpha^2_\psi} \sin (\Delta \Phi)$ and $\gamma_\psi = \sqrt{1-\alpha^2_\psi} \cos (\Delta \Phi)$. Decay matrices that depend on the parameters $\alpha$ and $\phi$ are used to handle the transformation of the spin operators in the hadronic two-body decays $D(B\to b \pi)$
\begin{align}
    \sigma^B_\mu \to \sum_{\nu=0}^3 a_{\mu,\nu}^D\sigma^b_\nu.
\end{align}
Combined, these tools let us express the full angular distribution for any baryon-antibaryon pair and its decay products in a very compact form. For a spin-1/2 baryon-antibaryon pair decaying in a single step, the angular distribution is given by
\begin{align}
    \mathcal{W}(\xi, \omega) = \sum_{\mu,\nu=0}^3 C_{\mu\nu}a^D_{\mu0}a^{\bar{D}}_{\nu0},
\end{align}
where $\xi$ is the set of helicity angles needed to completely describe an event and $\omega := (\alpha_\psi, \Delta \Phi, \alpha, \bar{\alpha})$ is the set of production and decay parameters. The formalism allows for a straightforward description of sequential decays trough multiplication by the corresponding decay matrices. Each additional matrix brings in further parameters and angular dependences.

\section{Experimental Methods}
With a total of $10^{10}$ and $3\times10^9$ events, respectively, the BESIII experiment has collected world-leading data samples at the $\jpsi$ and $\psi'$ resonances. The results presented in this contribution are mainly based on subsets consisting of $1.31\times10^9$ $J/\psi$ and $448\times10^6$ $\psi'$ events. Baryon-antibaryon pairs are reconstructed using two different strategies: double-tag (DT), where both the baryon and antibaryon are reconstructed, and single-tag (ST) where either the baryon or antibaryon is reconstructed and the missing mass is analyzed to ensure that the undetected particle is of the correct species. The former has the advantage that it yields complete information about the event, while the latter generally results in larger data samples. In both cases, all relevant helicity angles are measured and the production and decay parameters are determined through an unbinned maxmimum log-likelihood fit. 

\section{Recent Results from BESIII}

\subsection{The Reaction $e^+e^- \to \jpsi \to \Lambda \Lambdabar$}
In 2018, BESIII reported the first observation of polarized $\Lambda$ hyperons in decays of $\jpsi$ based on a data sample of $1.3\times10^9$ events~\cite{BESIII:2018cnd}. The non-zero polarization allowed for a simultaneous determination of the decay parameters of both $\Lambda$ and $\Lambdabar$ and a precise CP-test $A_{CP} =$ $-0.006\pm0.012\pm0.007$. Given that no CP-violation was observed, the average value was calculated in order to provide a more precise result $<\alpha> = \frac{\alpha - \bar{\alpha}}{2} =0.754\pm0.003\pm0.002$. At the time of publication, the value presented by BESIII was 17\% larger than the PDG average value that had been in place since the 1970s~\cite{ParticleDataGroup:2018}. Since then, the CLAS experiment has re-analysed its data on kaon photoproduction and found a value $\alpha_\Lambda = 0.721(6)(5)$~\cite{PhysRevLett.123.182301} that is likewise larger than the old PDG average, albeit not consistent with the BESIII result. The LHCb experiment has found that its data on $\Lambda_b \to \jpsi \Lambda$ favors the new BESIII result over the old PDG average~\cite{LHCb:2020iux}.

\begin{figure}[ht]
    \centering
    \makebox[\linewidth][c]{%
    \begin{overpic}[width=0.45\textwidth]{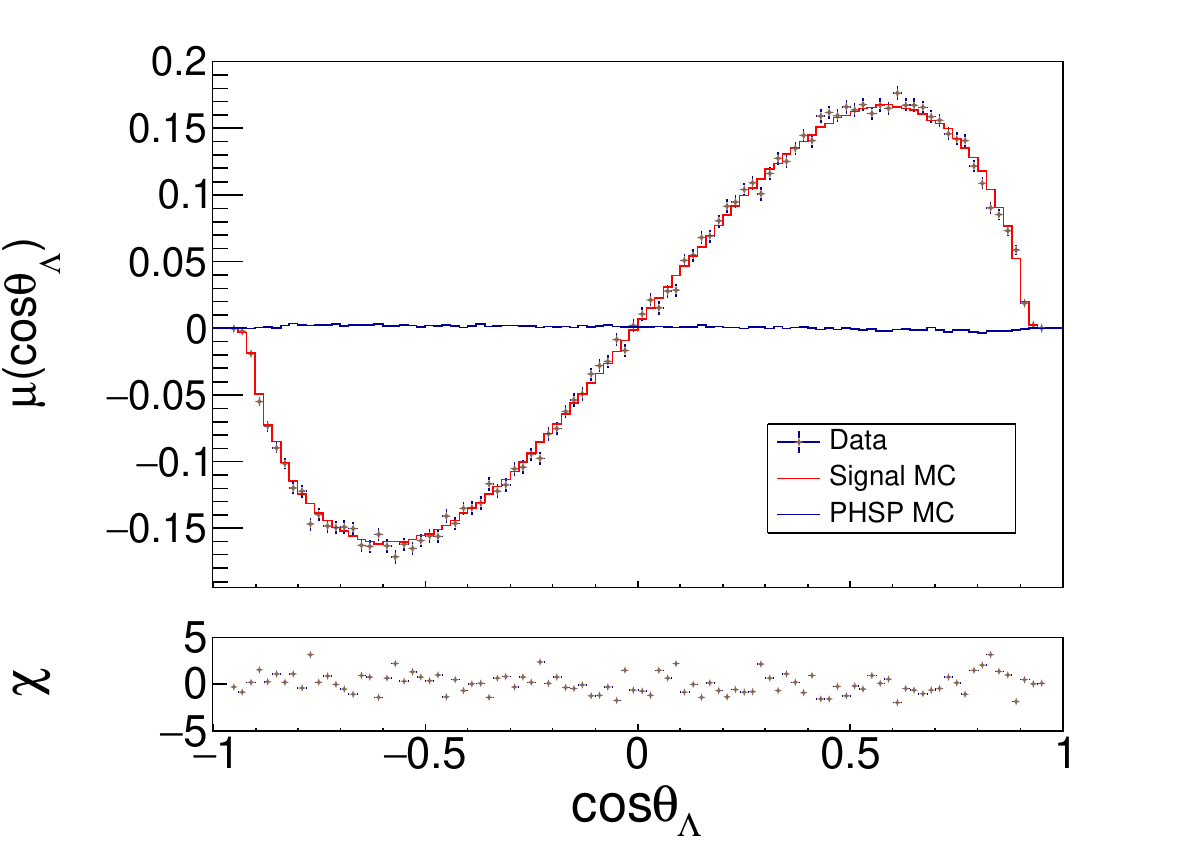}
    \put(80,62){\bf \large (a)}
    \end{overpic}
    \begin{overpic}[width=0.44\textwidth]{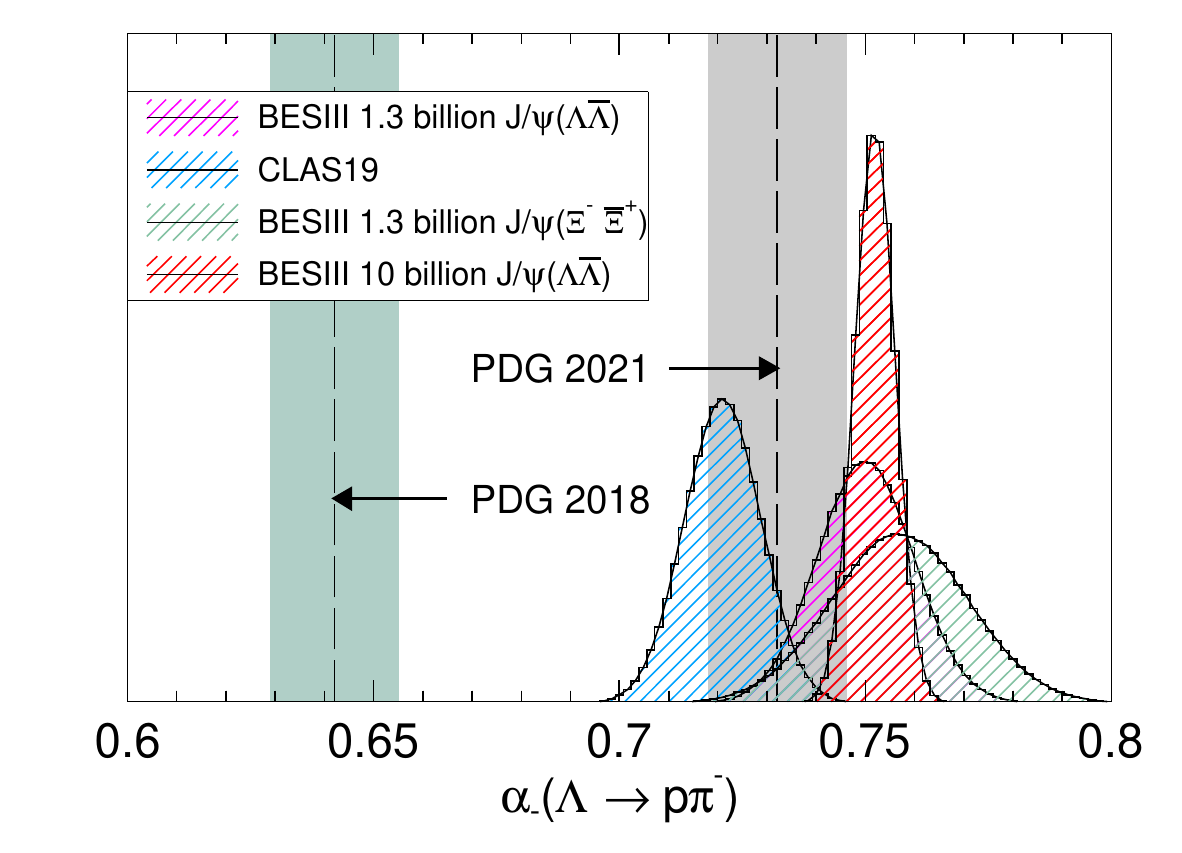}
    \put(80,62){\bf \large (b)}
    \end{overpic}}
    \caption{(a) Moment $\mu$ calculated using efficiency corrected data. Blue points with error bars represent data, the red line represents the fit results, and the blue line represent an unpolarized distribution. (b) Comparison of the recent measurements of the $\Lambda$ decay asymmetry parameter. The old PDG average is shown for reference.}
    \label{fig:LamLam}
\end{figure}

Recently, the BESIII experiment has performed a new analysis of the full dataset consisting of $10$ billion $\jpsi$ events~\cite{BESIII:2022qax}. The results are in good agreement with the previous measurement but with improved statistical precision $A_{CP} = -0.0025\pm0.0046\pm0.0012$. No CP-violation is observed, and the average decay asymmetry parameter is determined to be $<\alpha_> = 0.7542\pm0.0010\pm0.0024$. As a cross check of the fit result, the moment $\mu$ which is related to the polarization is calculated in 100 bins of $\cos \theta_\Lambda$, see Fig.~\ref{fig:LamLam}(a). Figure~\ref{fig:LamLam}(b) shows the current status for the $\Lambda$ decay asymmetry parameter.

\subsection{The Reaction $e^+e^- \to \Lambda \Lambdabar$}
The first complete measurement of the $\Lambda$ electromagnetic form factors was recently performed at a center-of-mass energy of 2.396 GeV using energy scan data collected at BESIII in 2015~\cite{BESIII:2019nep}. Fixing the values of the decay parameters to the values determined in the analysis of $\jpsi \to \Lambda \Lambdabar$, the relative phase between electric and magnetic form factors was determined for the first time for any baryon $\Delta \Phi = 37\pm12_{stat.}\pm6_{syst.}^\circ$. The ratio $R= |G_E/G_M| =$ $=0.96\pm0.14_{stat.}\pm0.02_{syst.}$ was also determined.

The reaction $e^+e^- \to \Lambda \Lambdabar$ was also studied at 3.773 GeV, just above the $\psi(3770)$ resonance. It was found that the branching fraction $\mathcal{B}(\psi(3770) \to \Lambda \Lambdabar)$ is more than ten times larger than has been previously assumed. This calls for a reevaluation of results on the $\Lambda$ electromagnetic form factors from the CLEO-c experiment near the $\psi(3770)$ resonance~\cite{Dobbs:2017hyd}. Through a full angular analysis, BESIII has determined the ratio $R_\psi = 0.48_{-0.35}^{+0.21}\pm0.03$ and relative phase $\Delta \Phi_{\psi} = 71_{-46}^{+66}\pm5^\circ$. Note that these values should be interpreted as effective values for the combination of the continuum and resonant contribution from the nearby $\psi(3770)$.

\subsection{The Reaction $e^+e^- \to  \jpsi/\psi' \to \Sigma \bar{\Sigma}$}
BESIII has furthermore observed spin polarization of $\Sigma^+$ in decays of both $J\psi$ and $\psi'$ for the first time~\cite{PhysRevLett.125.052004}. In both cases, a DT analysis is performed, and the non-zero polarization allows for the determination of the decay parameters $\alpha_{\Sigma^+}$ and $\alpha_{\bar{\Sigma}^-}$. These are in turn used for the first CP-test in decays of $\Sigma^+$: $\Sigma^+$ decays $A_{CP}^{\Sigma^+}=-0.004\pm0.037\pm0.010$. This result is to be compared with an SM prediction of $A_{CP}^{\Sigma^+} \sim 3.6\times10^{-6}$~\cite{Tandean:2003fr}.

\subsection{The Reaction  $e^+e^- \to  \jpsi \to \Xi^- \bar{\Xi}^+$}
Sequential decays of double strange hyperons allow for measurements of both the $\alpha$ and $\phi$ parameters and a direct extraction of the weak phase difference. A DT analysis of the reaction $J/\psi \to \Xi^-\bar{\Xi}^+ \to \Lambda \pi^- \bar{\Lambda}\pi^+ \to p \pi^- \pi^- \bar{p} \pi^+\pi^+$ based on $1.3\times 10^9$ $\jpsi$ events was recently performed by BESIII~\cite{BESIII:2021ypr}, finding a total of 73,244 signal events with an estimated background of $187\pm16$ events. In order to completely describe the polarization and entanglement from production through the two-step decay, nine helicity angles must be determined for each event. This allows for the determination of eight parameters $\omega_\Xi := (\alpha_\psi, \Delta \Phi, \alpha_\Xi, \phi_\Xi, \bar{\alpha}_\Xi, \bar{\phi}_\Xi, \alpha_\Lambda, \bar{\alpha}_\Lambda)$. The $\Xi$ produced in decays of $\jpsi$ are found to be polarized and this leads to the first measurement of the decay parameters of both $\Xi$ and $\bar{\Xi}$. Furthermore, this analysis gives an independent measurement of the decay parameters of $\Lambda$ and $\Lambdabar$. The average value $<\alpha_\Lambda> =$ $0.760\pm0.006\pm0.003$ is consistent with the result of the BESIII analysis of $J/\psi \to \Lambda \bar{\Lambda}$. In total, three independent tests of CP-symmetry are performed: $A_{CP}^\Xi=(6.0\pm13.4\pm5.6)\times10^{-3})$, $A_{CP}^\Lambda=$$(-3.7\pm11.7\pm9.0)\times 10^{-3})$, and $(\xi_P-\xi_S)_{\Xi^-\to \Lambda \pi^-} = (1.2\pm3.4\pm0.8) \times 10^{-2}$ rad. The latter is the first measurement of the weak phase difference for any baryon.

\subsection{The Reaction $e^+e^- \to \psi' \to \Omega \bar{\Omega}$}
\begin{figure}[ht]
    \centering
    \makebox[\linewidth][c]{%
    \begin{overpic}[width=0.4\textwidth]{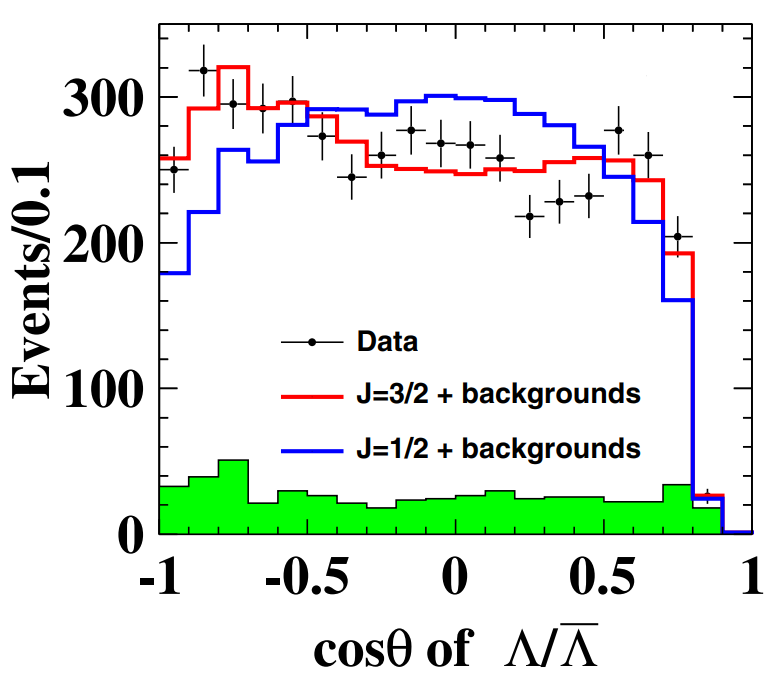}
    \put(86,72){\bf \large (a)}
    \end{overpic}
    \begin{overpic}[width=0.43\textwidth]{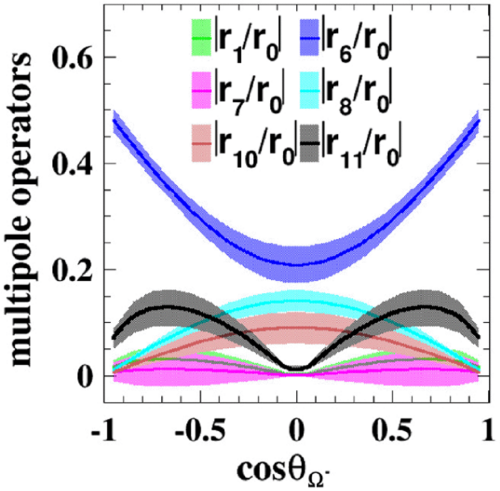}
    \put(84,79){\bf \large (b)}
    \end{overpic}}
    \caption{Comparison of experimental data with Monte Carlo simulations assuming $J=1/2$ and $J=3/2$ (a) and $\cos \theta_{\Omega^-}$ dependence of the $\Omega^-$ polarizations $r_i$ in $\psi'\to \Omega \bar{\Omega}$ (b).}
    \label{fig:OmegaSpinTest}
\end{figure}

The quark model spin of the $\Omega$ ($J=3/2$) has thus far not been tested model independently, but only under the assumption that the spins of $\Xi_c$ and $\Omega_c$ are $J=1/2$~\cite{BaBar:2006omx}. In a recent ST analysis of the process $\psi' \to \Omega^-\bar{\Omega}^+$ through measurements of the decay sequence $\Omega^- \to \Lambda K^-, \, \Lambda \to p\pi^-$ and its charge conjugate~\cite{PhysRevLett.126.092002}, BESIII has tested both the $J=1/2$ and $J=3/2$ hypotheses by fitting the corresponding formalisms to data. A comparison of the two hypotheses to data is shown in Fig.~\ref{fig:OmegaSpinTest}(a). The data clearly favor a spin of 3/2, confirming the quark model expectation model independently for the first time. Since its spin is 3/2, $\Omega$ can have vector ($r_1$), quadrupole ($r_6$,$r_7$,$r_8$), and octupole ($r_{10}$,$r_{11}$) polarizations. The values of the corresponding operators are determined in the fit and shown in Fig.~\ref{fig:OmegaSpinTest}(b).

\section{Outlook}
The BESIII experiment has now provided results on $J/\psi \to \Lambda \Lambdabar$ using the full dataset of $10^{10}$ $J/\psi$ events but for other channels, the statistical uncertainty on the CP-tests can still be improved by to a factor of about 2.7. In the future, more than $10^{12}$ $J/\psi$ events may be produced at planned Super Charm-Tau factories~\cite{Levichev_2018, Luo:IPAC2018-MOPML013}, and with a polarized electron beam these facilities could reach a level of precision where the SM predictions could be directly tested~\cite{Kupsc:2021bsu}.

\bibliography{ISMD}

\end{document}